\documentclass[preprint,nofootinbib,aps,showpacs,showkeys]{revtex4}

\usepackage{graphics}
\usepackage{amsfonts}
\usepackage{latexsym}

\newcommand{\lsim}[1]{
\setlength{\unitlength}{12pt}
\begin{picture}(1.4,1.)
\put(.7,-0.3){\makebox(0.0,1.)[t]{$<$}}
\put(.7,-0.3){\makebox(0.0,1.)[b]{$\sim$}}
\end{picture}#1}

\begin{document}

\title{Renormalization-group running cosmologies \\ - 
a scale-setting procedure}

\author{A. Babi\'c}
\email{ababic@thphys.irb.hr}
\author{B.Guberina}
\email{guberina@thphys.irb.hr}
\author{R. Horvat}
\email{horvat@lei3.irb.hr}
\author{H. \v Stefan\v ci\'c \footnote{Presently at the Departament d' Estructura i Constituents de la Mat\`eria, Universitat de Barcelona, Av. Diagonal 647, 08028 Barcelona, Catalonia, Spain. On leave of absence from
the Theoretical Physics Division, Rudjer Bo\v{s}kovi\'{c}
Institute, Zagreb, Croatia.}}
\email{shrvoje@thphys.irb.hr}

\affiliation{Rudjer Bo\v{s}kovi\'{c} Institute,
   P.O.Box 180, HR-10002 Zagreb, Croatia}


\begin{abstract} 
For cosmologies including scale dependence of both the cosmological 
and the gravitational constant, an additional consistency condition dictated by 
the Bianchi identities emerges, even if the energy-momentum tensor of ordinary
matter stays individually conserved. For renormalization-group (RG) approaches 
it is shown that such a consistency relation ineluctably fixes the RG scale 
(which may have an explicit as well as an implicit time dependence), provided 
that the solutions of the RG equation for both quantities are known. Hence,
contrary to the procedures employed in the recent literature, we argue that
there is no more freedom in identification of the RG scale in terms of the
cosmic time in such cosmologies. We carefully set the RG scale for the RG
evolution phrased in a quantum gravity framework based on the hypothetical
existence of an infrared (IR) fixed point, for the perturbative regime within
the same framework, as well as for an evolution within quantum field theory
(QFT) in a curved background. In the latter case, the implications of the
scale setting for the particle spectrum are also briefly discussed.
\end{abstract}


\noindent
\pacs{04.62.+v; 04.60.-m; 98.80.-k; 98.80.Es}
\keywords{Cosmological constant; Cosmology; Renormalization-group equation;
Scale setting}

\maketitle

Recently, indisputable evidence has been mounting to suggest that the
expansion of our universe is accelerating owing to the nonvanishing value of
unclustered dark energy with negative pressure, see \cite{1}. The crucial
evidence for the existence of dark energy [or the cosmological constant (CC)]
relies on the CMB observations \cite{2} which strongly support a spatially
flat universe as predicted by inflationary models. By combinations of all data
a current picture of the universe emerges, in which about 2/3 of the critical 
energy density of the present universe is made up by a background dark energy 
with the parameter of the equation of state $-1.38 \leq w \leq -0.82$ at $95\%
$ confidence level \cite{3}. Pressed by these data, theorists now need explain 
not only why the CC is small, but also why dark-energy domination over ordinary 
matter density has occurred for redshifts $z \lsim 1$ (the coincidence problem).

Although models with a truly static CC fit these data well, they have two
additional drawbacks (besides the coincidence problem): (1) they cannot
theoretically explain why the CC is today small but nonvanishing; (2) they
have a problem to ensure a phase of inflation, an epoch in the early universe 
when the CC dominated other forms of energy density. Assessing the possibility 
to have dynamical dark energy, rolling scalar field models (quintessence 
fields) \cite{35} with generic attractor properties \cite{4} that make the 
present dark energy density insensitive to the broad range of unknown initial 
conditions, have been aimed at dealing with the coincidence problem. Still, a 
quintessence potential has to be fine-tuned to yield the present ratio of 
ordinary matter to quintessence energy density, at the same time allowing a 
phase dominated by matter so that structure can form; therefore these models 
cannot address the coincidence problem. In addition, such models may have 
difficulties in achieving the current quintessence equation of state with its 
parameter below -0.8 \cite{5}. Although extended models with spatially 
homogeneous light scalar fields based on a nonlinear kinetic energy 
(k-essence) \cite{6} have been put forward, it seems that today a trustworthy 
solution to the coincidence problem, beyond invoking an anthropic principle 
\cite{7}, is still lacking. Recently, motivated by the observational data, 
models of dark energy with the supernegative equation of state ($w < -1$) have 
been introduced \cite{Cald}. This form of dark energy, named phantom energy, 
leads to many interesting phenomena, the most striking being the possibility of 
divergence of the scale factor in finite time, the so-called ``Big Rip" effect 
\cite{Cald2}.

Another class of variable dark-energy models which could successfully mimic 
quintessence models and may also shed some light on the coincidence problem, 
have been put forward recently. They are based on the observation \cite{8, 9}
that even a ``true" CC (with the equation of state being precisely -1) cannot 
be fixed to any definite constant (including zero) owing to the 
renormalization-group (RG) running effects. In \cite{9}, the variation of the 
CC arises solely from particle field fluctuations, without introducing any
quintessence-like scalar fields. Particle contributions to the RG  running of 
$\Lambda $ due to vacuum fluctuations of massive fields have been properly 
derived in \cite{10}, with a somewhat unexpected outcome that more massive 
fields do play a dominant role in the running at any scale. In the model 
\cite{11}, the RG running is due to non-perturbative quantum gravity effects 
and the hypothesis of the existence of an IR attractive RG fixed point. Both 
models \cite{9,11} also promote the gravitational constant to a RG running 
quantity \footnote{For interesting work on the scale dependent gravitational coupling see \cite{bertolami}.}, with a prominent scaling behavior found in \cite{11}. In both models 
the presence of quintessence-like scalar fields is redundant and not required 
for consistency with observational data.          

It should be noted that in the above RG running models the amount of running
will depend not only on the parameters of an underlying physical theory, but 
also on the characteristic RG scale, which must be correctly identified. 
Several different scenarios for RG-scale adoption have been contemplated in the
literature. In \cite{12} the RG scale was identified with the Hubble parameter,
which for the present time is $H_0 \sim  10^{-33} \;\mbox{\rm eV}$. This leads
to extremely slow running of the CC and the gravitational constant. Another
choice, given by the fourth root of the total energy density \cite{9, 10},
produces much faster running of the CC. The advantage of such a choice is that 
it entails a direct association with particle momenta (and therefore with the 
temperature of interacting particle species) in the past radiation epoch. 

In the models \cite{8,85,11} the RG scale (cutoff) was identified with the 
inverse of the cosmological time, which is essentially equivalent to the 
previous case of the Hubble parameter for cosmologies with $a \propto t^n $. 
Another choice for the relevant cutoff with the implicit time dependence in the 
form of the inverse scale factor was also analyzed \cite{85}. With the above 
choices, it was found  that no consistent solution existed in the former case 
for curved universes (aside for the radiation-dominated era) and even for flat 
universes in the latter case. In the present paper, we argue that the 
consistency relation as dictated by the Bianchi identity, which relates the 
time dependencies of the CC, the gravitational constant and the ordinary matter 
density, does unambiguously set the RG scale once the RG solutions are known. 
Hence, any conclusion about the consistency of cosmological solutions with an 
arbitrary choice of the RG scale may be quite misleading. Our scale-setting 
procedure works well for both flat and curved universes as well as for the 
matter background with an arbitrary equation of state. We note that if the 
ordinary energy-momentum tensor is not  separately conserved (for instance, 
owing to the interaction between matter and the CC, which causes a transfer of 
energy from matter to the CC or vice versa), then the consistency relation 
becomes more complicated and the scale-setting procedure is no longer 
straightforward. In the following, we explain the scale-setting procedure in 
detail and apply it to several relevant  RG solutions obtained in quantum 
gravity and in QFT in a curved background.

\section{{\bf Scale-setting procedure: importance and limitations}}

\label{sec2}

In many cosmological approaches based on the renormalization-group running of
the fundamental quantities such as $\rho_{\Lambda}$ and $G$, the fundamental
procedure of the underlying quantum (field) theory specifies the running of the
aforementioned quantities in terms of the appropriate running scale. From the
viewpoint of quantum field theory, the running scale has a natural 
interpretation in the form of the scaling factor of external momenta or the 
energy cutoff. The formulation of the consistent cosmological model comprising 
running quantities requires translation of the running in terms of the running 
scale to the evolution of these quantities in terms of cosmological variables. 
This identification of the running scale with a specific function of 
cosmological variables is a crucial step in all approaches relying on the RGE 
approach for the dynamics of quantities such as $\rho_{\Lambda}$ and $G$. All 
efforts that have been undertaken in this direction so far have concentrated on 
the argumentation in favor of some seemingly ``natural'' choices for the RGE 
scale, 
without a proper procedure of their derivation. 

What is meant by ``natural'' choices is to a great extent
a matter of taste and depends
strongly on the outcome one wishes to achieve. Thus, one should not be
surprised that the literature contains choices for  the
RG scale which, at present, differ up to 30 orders of magnitude, see,
e.g.,
Refs.[13, 14, 16]. Needless to say, with different choices for the RG scale
one  automatically selects  different cosmologies, thus making RG
approaches devoid of any firm and generic prediction.

In addition to noting that even a phenomenological setting of the RG
scale
is not an obvious matter in cosmological setups, one also finds  for
both RG cosmologies discussed in this paper (for detailed discussions, see the
subsequent chapters below) that the situation with the scale setting is
intrinsically not free from ambiguities. For instance, for the RG evolution
in a conventional field-theoretical model in the classical curved
background (see chapter 6), one derives the beta functions from the loop
expansion. The main ambiguity comes from the fact that the beta functions
are strongly dependent on the renormalization scheme. For example, in the 
MS-scheme
the RG shows how the various parameters depend on the RG scale $\mu $. On
the other hand, in the mass-dependent schemes there is no $\mu $, and one
directly gains the dependence on the external signal (like the typical
energy of interaction). Aside from the UV limit where both schemes are
proved to be equivalent, this identification fails, especially for
massive fields. As we deal here  with vacuum graphs with no external legs,
not only the physical sense of $\mu $ becomes more troublesome, but also
decoupling of heavy-particle species remains ambiguous in both schemes. In
the
RG approach in quantum gravity (see chapters 3 and 4), which is essentially
nonperturbative, one is shown how a given observable is related to the IR
cutoff in the theory. Hence, the role of $\mu $ is undertaken by the IR
cutoff $k$. The cosmological interpretation of $k$ is less ambiguous now
because
it should approximately match the size of the system; one can envisage
several
candidates like the Hubble distance, the particle horizon, or the event
horizon. Still, the intrinsic interpretation of $k$ is still ambiguous as any link
to the external signal is missing.

Despite the phenomenologically 
interesting results and more or less strong physical motivation behind some 
choices for the RGE scale, the fact remains that the RGE scale-setting 
procedure is largely arbitrary. A naturally arising question is whether some 
physical argument might remove this apparent arbitrariness and introduce a 
procedure that would set the RGE scale automatically in terms of cosmological 
variables. This paper is centered around such a procedure for a broad class of 
models based on the RGE approach.

The class of RGE-based cosmological models is characterized by the following
properties. The only running quantities taken into consideration are the
cosmological term $\Lambda$ (or equivalently the cosmological term energy
density $\rho_{\Lambda}= \Lambda/(8 \pi G)$) and the gravitational coupling
constant $G$. These running quantities depend on a single running scale,
generally denoted by $\mu$. Other components of the content of the universe
(such as nonrelativistic matter or radiation) evolve in a standard way, i.e.
there is no exchange of energy-momentum between these components and the 
dynamical cosmological term. For the sake of simplicity and clarity of 
exposition, in all considerations put forward in this paper we assume that 
there exists only one additional matter component described by the equation of 
state $p_{m}= w \rho_{m}$, where $w$ is constant. The inclusion of additional 
matter components and/or more general equations of state is straightforward. 

The physical principle behind the scale-setting procedure is conceptually 
simple. The RGE improved Einstein equation for these cosmologies acquires the 
form
\begin{equation}
\label{eq:Ein}
R^{\alpha\beta} - \frac{1}{2} g^{\alpha\beta} R = -8 \pi G(\mu)(
T_{m}^{\alpha\beta} + T_{\Lambda}^{\alpha\beta}(\mu)) \, ,
\end{equation}
where $R^{\alpha\beta}$ and $R$ denote the Ricci tensor and scalar respectively,
while $T_{m}^{\alpha\beta}$ and $T_{\Lambda}^{\alpha\beta}(\mu) = \rho_{\Lambda}
(\mu) g^{\alpha\beta}$ stand for matter and cosmological term energy-momentum 
tensors, respectively. The only physical requirement is that equation 
(\ref{eq:Ein}) must maintain its general covariance with the quantities $G$ and $T_{\Lambda}^{\alpha\beta}$ having the $\mu$ (and implicitly $t$) dependence.

This requirement translates into the following equation relating $G$, $\rho_{
\Lambda}$ and $\rho_{m}$:
\begin{equation}
\label{eq:evol}
\dot{G}(\mu)(\rho_{m}+\rho_{\Lambda}(\mu)) + G(\mu)\dot{\rho_{\Lambda}}(\mu)
=0\, .
\end{equation}
Here dots denote time derivatives. Assuming the nonvanishing $\dot{\mu}$ 
throughout the evolution of the universe, we arrive at the equation
\begin{equation}
\label{eq:muset}
\rho_{m} = -\rho_{\Lambda}(\mu) - 
G(\mu) \left( \frac{d \rho_{\Lambda}(\mu)}{d \mu}
\right) \left( \frac{d G(\mu)}{d \mu} \right)^{-1} \, .
\end{equation}
On the left-hand side of the above equation we have a function of the scale factor $a$, 
while on the right-hand side we have a function of $\mu$, i.e. we 
have an expression of the form $\rho_{m}=f(\mu)$. The inversion of this 
expression in principle gives the scale $\mu$ in terms of $\rho_{m}$ in the 
form $\mu=f^{-1}(\rho_{m})$. In the cases in which the inversion of the
function $f$ gives multiple possibilities for the choice of $\mu$, we assume
that one of them can be selected on some physical grounds (such as the
positivity of the scale in the case of $f(x)=x^2$).

Once the procedure of the scale setting is completed, we formally have the
quantities $\rho_{\Lambda}$ and $G$ as functions of $\rho_{m}$ \footnote{
However, it is necessary to stress that these dependences are not fundamental, 
but derived in a scale-setting procedure.}. The matter energy density has the 
usual scaling behavior with the scale factor
\begin{equation}
\label{eq:scalmatt}
\rho_{m}=\rho_{m,0}\left( \frac{a}{a_{0}} \right)^{-3(1+w)} \, .
\end{equation}
The set of equations necessary for the description of the evolution of the
universe is completed by the Friedmann equation
\begin{equation}
\label{eq:Friedmann}
\left(\frac{\dot{a}}{a} \right)^{2} + \frac{K}{a^{2}} = \frac{8 \pi}{3}
G (\rho_{m}+\rho_{\Lambda})\, .
\end{equation} 

In the following sections we present several examples of the functioning of the 
procedure and point out some physically interesting effects in the evolution of 
the cosmological models based on RGE and our scale-setting procedure.  

Before doing this, we make some comments on the character of our
scale-setting procedure  as well as on cases where it
is not applicable. First of all, it is  clearly seen that our procedure
lacks the first-principle connection to quantum gravity considerations,
and therefore should be considered as purely phenomenological.
Concerning our basic Eq. (2), we have assumed that $\rho_m $
is intrinsically independent of $\mu $. \footnote{We can safely ignore weak
logarithmic running of masses contained in $\rho_m $.} As seen from (2), as
long as $\rho_m $ retains its canonical form, the scale is univocally fixed
and  does not even implicitly depend on $\mu $ in (2). The situation
changes when, for instance, an interaction between matter and the CC and/or $G$ is turned on. In this case,  Eq. (2) generalizes to
\begin{equation}
\dot{G}(\rho_{\Lambda } + \rho_m ) + G \dot{\rho }_{\Lambda } +
G (\dot{\rho }_{m} + 3H\rho_m (1 + w ))  = 0  \;.
\end{equation}
We see from the above equation for the interacting matter that any
deviation from the canonical case (4) depends decisively on $\mu $. Hence,
one is not able to fix the scale before specifying such
interactions {\it  a priori}. Also, the same conclusion is reached 
for other two extra cases
derived from (6): (i) for the constant $G$ and the running $\Lambda $, and 
(ii) for the constant
(or zero) $\Lambda $ and the running $G$. Thus, one is restricted to the
cosmological setup defined by (2) for the full applicability of our
scale-setting procedure.

As a first step in the next section, we shortly review the nonperturbative
quantum gravity as derived by Reuter et~al. \cite{8,85} and apply the scale
setting procedure for the running $\Lambda (k)$ and $G(k)$ in the perturbative
regime, which interpolates the behavior of $\Lambda (k)$ and $G(k)$ between the
UV fixed point and the infrared limit.

\section{Exact renormalization-group approach in quantum gravity}

\label{sec2.5}

In the quantum-gravity approach of \cite{QG1, QG2} the metric is the 
fundamental dynamical variable. It is possible to construct a scale-dependent 
gravitational action $\Gamma_k [g_{\mu\nu}]$ and derive the appropriate 
evolution equation. The properties of the constructed ``effective average 
action", especially a built-in infrared cutoff, are welcome in the case of 
quantum gravity since it is possible to study low-momentum behavior in a 
nonperturbative way. In such a way, quantum gravity can be used to describe 
gravitational phenomena at very large distances. 

A systematic study of the application of the exact renormalization-group 
approach to quantum gravity\footnote{Similar problems were also treated in \cite{Odintsov,Litim}.} was done by Reuter \cite{8}. There appears the 
effective average action $\Gamma_k [g_{\mu\nu}]$, which is basically a 
Wilsonian coarse-grained free energy \cite{QG2, reuwett}.  This action
depends on metrics and the momentum scale $k$  interpreted as an
infrared cutoff in the following way.
If one has a 
physical system with a size $L$, then the mass parameter $k \sim 1/L$ defines 
an infrared cutoff.
This is the fundamental step in deriving the average
effective action in the Wilson-Kadanoff formulation of the effective action.
The main difference between the effective average action 
$\Gamma _k [g_{\mu\nu}]$ and the ordinary effective action $\Gamma [g_{\mu\nu}]$ is in the fact that the path integral which defines $\Gamma _k$ integrates only the quantum fluctuations with the momenta $p^2 > k^2$, thus describing  
the dynamics of the metrics averaged over the volume with size
$(k^{-1})^3$. The derived effective field theory is valid near the scale $k$.  For any scale $k$, there 
is an $\Gamma_k$ which is an effective field theory at that scale. This means 
that all gravitational phenomena are correctly described at tree level by 
$\Gamma_k$, including the loop effects with $p^2\simeq k^2$. Quantum 
fluctuations for $p^2 > k^2$ are integrated out, and large-distance metric 
fluctuations, $p^2 < k^2$, are not included. Of course, in the limit  
$k\rightarrow 0$, the infrared cutoff ``disappears" and the original effective 
action $\Gamma$ is recovered. On the other hand, the Einstein-Hilbert action 
can be interpreted as fundamental theory in the limit $k\rightarrow \infty$.

The determination of the infrared cutoff (i. e., the scale $k$)
is by no means trivial in reality. It is rather simple in a massless theory such as massless QED, where the inverse of a distance is the only mass scale
present in the theory. However, a real situation one
encounters is the variety of mass
scales and the proper identification, if any, is not trivial at all. The 
correct way to proceed would be to study the RG flow of the effective action
$\Gamma _k [g_{\mu\nu}]$ and make the identification of the infrared cutoff
by inspecting 
the RG evolution. Once the infrared cutoff is identified, one should solve the
 Bianchi identities and the conservation laws for matter \cite{Martin}. 
On the other hand, one can start vice versa, i. e., use the Bianchi identities,
etc., to fix a scale and look whether it is possible to give 
a physical interpretation to the scale.

It is clear that there should exist a certain physical mechanism which 
effectively stops the running in the infrared. Obviously, it is a function of 
all possible scales that appear in a certain case and/or characterize a 
physical system under consideration --- it may include particle momenta, field 
strengths, the curvature of spacetime, etc. However, looking at the global 
system described by Einstein equations, the conditions of homogeneity and 
isotropy lead to a possible natural choice of $k$ to be proportional to the 
cosmological time $t$, $k=k(t)$. If the dependence $k(t)$ is known, the $G(k)$ 
and $\Lambda (k)$ become the functions $G(t)$ and $\Lambda (t)$. 

The choice $k\propto \frac{1}{t}$ seems to be a plausible one. The cosmological 
time $t$ describes the temporal distance between some event-point $P$ and the 
big-bang. For the effective field theory $\Gamma_k$ to be valid at $t$, one 
need not integrate quantum fluctuations with momenta smaller than $\frac{1}{t}$, 
$p^2\ll \frac{1}{t^2}$. Namely, at the time $t$, the fluctuations with 
frequencies smaller than $1/t$ should not play any role yet, and the meaning of 
the infrared cutoff becomes obvious.

Nonperturbative solutions of the RG equations are obtained by truncation of the 
original infinite dimensional space of all action functionals to some specific 
finite dimensional space which appears relevant to a given physical problem. 
The usual truncation is the reduction to the Einstein-Hilbert action. This 
Ansatz leads to the coupled system \cite{8,85} of equations for $G(k)$ and 
$\Lambda (k)$. For the case $\Lambda < k^2$, it is simple and easy to solve. 
It leads to two attractive fixed points $g_*$, $\beta (g_*)=0$.

The ultraviolet (non-gaussian) fixed point is
\begin{equation}
g_*^{\mathrm{UV}}=\frac{1}{\omega '},
\end{equation}
where $\omega '$ is the number which is calculated, and an infrared (gaussian) 
fixed point at $g_*^{\mathrm{IR}}=0$. All trajectories which are attracted 
towards $g_*^{\mathrm{IR}}=0$ for $k=0$ and towards $g_*^{\mathrm{UV}}$ for 
$k\rightarrow \infty$ lead to the following form of $G (k)$:
\begin{equation}
G(k)=\frac{G_0}{1+\omega G_0 k^2} \, ,
\end{equation}
which for small $k$ reads 
\begin{equation}
G(k)=G_0-\omega G_0^2 k^2 + {\mathcal{ O}}(k^4)	\, .		\label{eq:g2}
\end{equation}
%
%

Since for the Einstein-Hilbert truncation $G(k)$ does not run between the scale 
where it is experimentally determined to be the Newton constant $G_{\mathrm{N}
}$, and a cosmological scale for which $k\sim 0$, $G_0$ can be safely identified 
as $G_0\equiv G(k=0)=G_{\mathrm{N}}$.

For $k^2\gg G_0^{-1}$, the fixed-point behavior sets in and $G(k)$ becomes
\begin{equation}
G(k)\sim \frac{1}{\omega k^2} \, ,
\end{equation}
in accordance with the asymptotic running predicted by Polyakov \cite{polyakov}.
The cosmological constant $\Lambda (k)$ then reads
\begin{equation}
\Lambda (k)=\Lambda_0+\nu G_0 k^4\left[
1+{\mathcal{O}}(G_0 k^2)\right] \, .							\label{eq:g4}
\end{equation}
%
%

Equations (\ref{eq:g2}) and (\ref{eq:g4}) define the so-called perturbative 
regime of the infrared cutoff $k$. The solutions (\ref{eq:g2}) and 
(\ref{eq:g4}) are basically expansions in the dimensionless ratio $(k/M_{Pl
})^2$. By inspection one sees that renormalization effects are important only 
for the infrared cutoff $k$ approaching $M_{Pl}$. Equation (\ref{eq:g2}) shows 
that increasing $k$ leads to decreasing $G (k)$ --- a first sign of asymptotic 
freedom of pure quantum gravity \cite{8}.

If we apply our scale-setting procedure to the quantum gravity we reviewed, 
i.e. using (\ref{eq:g2}) and (\ref{eq:g4}) for $G(k)$ and $\Lambda (k)$, and 
the expression (\ref{eq:muset}) for $k$, we obtain 
\begin{equation}
k=\sqrt{\frac{4\pi\omega}{\nu}\rho_{m}G_0},
\end{equation}
and
\begin{equation}
k_0\sim\sqrt{\rho_{m, 0}G_0}\sim H_0
\end{equation}
for the ``present" value of the infrared cutoff $k_0$.
%
%

\section{Quantum-gravity model with an IR fixed point}

\label{sec3}

In the formalism of quantum gravity introduced in \cite{8,85}, it is possible
to set up renormalization-group equations for the cosmologically relevant
quantities $\Lambda$ and $G$. From the phenomenological point of view, the
assumption of the existence of the IR fixed point in the RGE flow \cite{11} is 
of special interest (for related work see also \cite{Litim}).\footnote{
In the effective framework of Ref. \cite{8, 85} based on the
Einstein-Hilbert approximation, an analysis of IR effects leading to the
existence of the IR fixed point in the RGE flow is not possible. It would
require truncations beyond the Einstein-Hilbert approximation, containing
nonlocal invariants, for instance. Hence, the existence of the IR fixed
point  was hypothesized in \cite{85}.}
 
In this setting, the infrared cutoff $k$ plays the role of
the general RGE scale $\mu$. In the infrared limit, the running of $\Lambda$ 
and $G$ can be expressed as
\begin{equation}
\label{eq:LamIR}
\Lambda(k) = \lambda_{*} k^2 \, ,
\end{equation}
\begin{equation}
\label{eq:GIR}
G(k) = \frac{g_{*}}{ k^2} \, ,
\end{equation}
where $\lambda_{*}$ and $g_{*}$ are constants and $k$ is the infrared cutoff.

The application of the procedure explained in section \ref{sec2} to the running
parameters (\ref{eq:LamIR}) and (\ref{eq:GIR}) results in expressing the scale
$k$ in terms of cosmological quantities (and implicitly of time). Insertion of 
(\ref{eq:LamIR}) and (\ref{eq:GIR}) into (\ref{eq:muset}) finally gives
\begin{equation}
\label{eq:IRset}
k = \left( \frac{8 \pi g_{*}}{\lambda_{*}} \rho_{m} \right)^{1/4} \, .
\end{equation}
It is important to note that the result of this scale-setting procedure is not 
explicitly dependent on the topology $K$ of the universe. One may, however, 
argue that there exists a certain level of implicit dependence since the 
expansion of the universe (and therefore the dependence of $\rho_{m}$ on time) 
depends on $K$. Moreover, the scale-setting procedure of this paper 
unambiguously identifies the scale $k$ in the model of quantum gravity with the 
IR fixed point.

The result (\ref{eq:IRset}) leads to the following general result:
\begin{equation}
\label{eq:LamrelG}
\Lambda = 8 \pi G \rho_{m} = (8 \pi \lambda_{*} g_{*} \rho_{m})^{1/2} \, .
\end{equation}

Combining the expressions (\ref{eq:scalmatt}), (\ref{eq:Friedmann}) and 
(\ref{eq:LamrelG}) allows the dependence of the scale factor on time. It is
convenient to introduce the parameters $\rho_{c} = 3 H^{2}/(8 \pi G)$, 
$\Omega_{\Lambda} = \Lambda/(8 \pi G \rho_{c})$, $\Omega_{m}=\rho_{m}/\rho_{c}$
and $\Omega_{K}=-K/(H^2 a^2)$. In terms of these parameters equation 
(\ref{eq:LamrelG}) can be written as
\begin{equation}
\label{eq:Omegas}
\Omega_{\Lambda} = \Omega_{m} \, .
\end{equation}

Using the relation $\Omega_{\Lambda} + \Omega_{m} + \Omega_{K} = 1$ we obtain 
an implicit expression for the scale factor of the universe:
\begin{equation}
\label{eq:aodt}
H_{0} (t-t') = \int_{a'/a_{0}}^{a/a_{0}} [\Omega_{K}^{0} + (1-\Omega_{K}^{0})
u^{(1-3w)/2} ]^{-1/2} \, d u \, .
\end{equation}
%
Throughout this paper the subscript or the superscript $0$ refers to the present
epoch of the evolution of the universe. From the expression (\ref{eq:aodt}) it
is evident that the expansion of the universe depends on two parameters: $w$ and
$\Omega_{K}^{0}$. The general solution of (\ref{eq:aodt}) can be given in terms 
of the confluent hypergeometric functions. However, it is far more instructive 
to consider the form that the solutions acquire for some special choices of 
these parameters. We concentrate on three interesting cases in the limit 
$t' \rightarrow 0$ ($a' \rightarrow 0$) to make a comparison with the results 
of \cite{11}:

{\bf i) $\mathbf{\Omega_{K}^{0}=0}$, $\mathbf{w}$ arbitrary.} For the flat 
universe one obtains
\begin{equation}
\label{eq:flat}
\frac{a}{a_{0}}=\left( \frac{3}{4} (1+w) H_{0} t \right)^{\frac{4}{3(1+w)}} \, ,
\end{equation}
which leads to the following expression for the scale $k$:
\begin{equation}
\label{eq:kflat}
k = \frac{4}{3(1+w)H_{0}} \left( \frac{8 \pi g_{*}}{\lambda_{*}} \rho_{m,0} 
\right)^{1/4} \frac{1}{t} = \frac{1}{1+w} \left( \frac{8}{3 \lambda_{*}} 
\right)^{1/2} \frac{1}{t} \, .
\end{equation}
The results obtained above are in complete agreement with the results of 
\cite{11} for the flat case. However, in the case of \cite{11}, the choice
$k=\xi/t$ was introduced on phenomenological grounds and it was found that
consistency was achieved in an otherwise overdetermined set of equations. In 
our case, the setting of the scale $k$ is a result of the systematic procedure 
and confirms the choice of \cite{11}. 

{\bf ii) $\mathbf{w=1/3}$, $\mathbf{\Omega_{K}^{0}}$ arbitrary.} In this 
cosmological model the universe of arbitrary curvature has a simple evolution 
law for the scale factor
\begin{equation}
\label{eq:wrad}
\frac{a}{a_{0}}= H_{0} t \, ,
\end{equation}  
which allows us to express $k$ as
\begin{equation}
\label{eq:krad}
k = \frac{1}{H_{0}} \left( \frac{8 \pi g_{*}}{\lambda_{*}} \rho_{m,0} 
\right)^{1/4} \frac{1}{t}  \, .
\end{equation}
The result displayed above further explains why the Ansatz $k=\xi/t$ functions
well also for cosmologies of any curvature including the radiation component 
only. We have found agreement with \cite{11} in this case as well.

\begin{figure}
\centerline{\resizebox{1.0\textwidth}{!}{\includegraphics{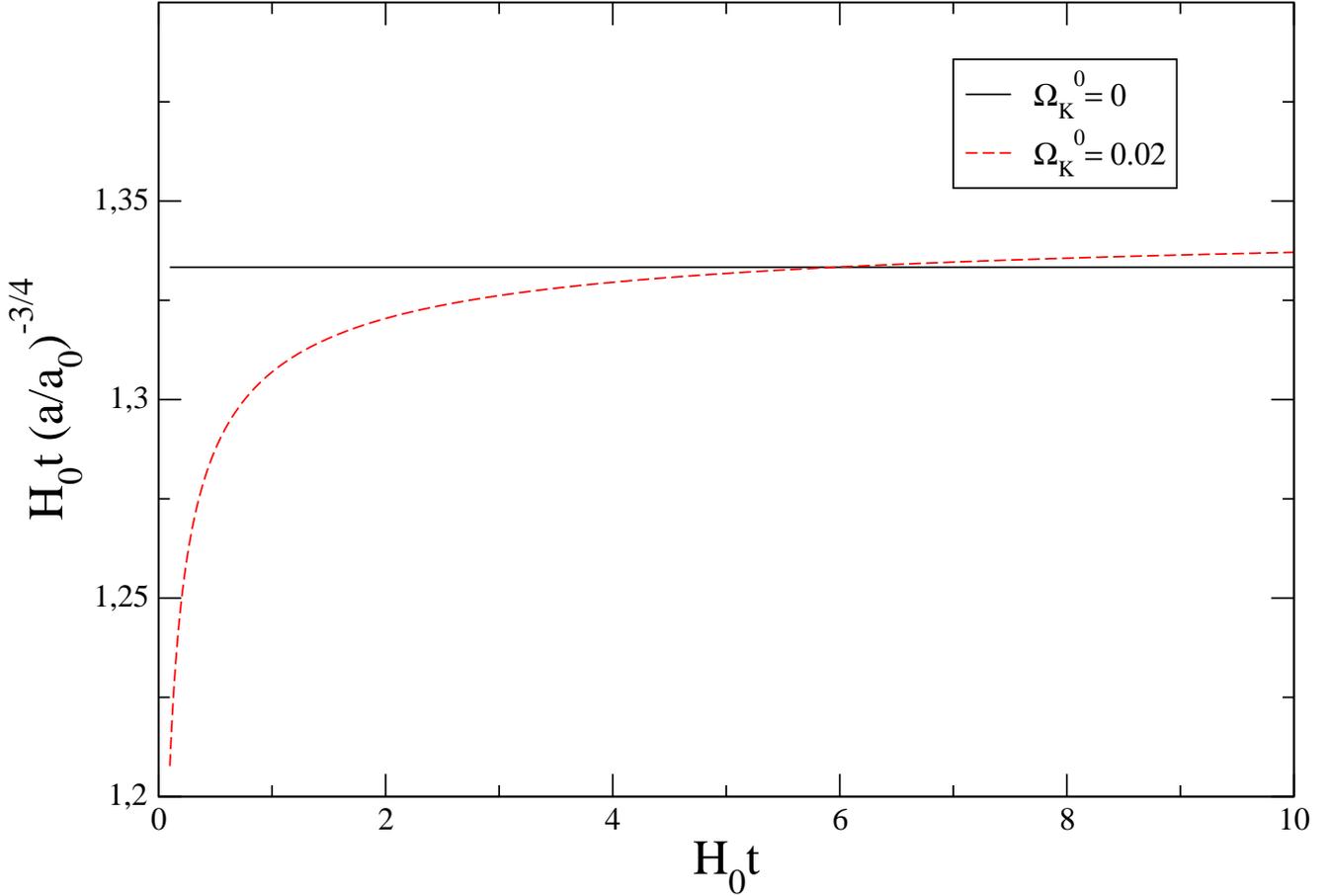}}}
\caption{\label{fig:1} The product of the cosmic time $t$ and $(a/a_{0})^{-3/4}$
versus $t$ is given for the flat and the open universe with $\Omega_{K}^{0}=
0.02$. The comparison of graphs shows that for the curved universe the choice 
of the scale differs from $k \sim 1/t$.}
\end{figure}

{\bf iii) $\mathbf{w=0}$, $\mathbf{\Omega_{K}^{0} \ge 0}$.} The universe 
containing nonrelativistic matter only and having arbitrary curvature has the 
law of evolution of the scale factor given by the following expression:
\begin{eqnarray}
\label{eq:nonrel}
H_{0} t &=& \frac{4}{3(1-\Omega_{K}^{0})^{2}} \left[
\left( \Omega_{K}^{0}+(1-\Omega_{K}^{0})\left( \frac{a}{a_{0}} 
\right)^{1/2} \right)^{1/2} \left(  (1-\Omega_{K}^{0})
\left( \frac{a}{a_{0}}\right)^{1/2} - 2 \Omega_{K}^{0} \right) 
\right. \nonumber \\
&+& 2 (
\left. \Omega_{K}^{0})^{3/2} \right] \, .
\end{eqnarray}
From the expression given above it is possible to obtain the scale factor $a$ 
as a function of time. In this case, the scale $k$ can no longer be expressed 
in the form of Ansatz $k = \xi/t$. However, it is still possible to uniquely 
define the scale $k$ by choosing the only real solution of the cubic equation 
(\ref{eq:nonrel}). The product $t (a/a_{0})^{-3/4}$ (since $k \sim 
(a/a_{0})^{-3/4}$) as a function of time is displayed in Fig. \ref{fig:1}. It 
is clear from the figure that the choice $k \sim 1/t$ is adequate for the flat
universe, while for the curved one the scale may differ from the $\xi/t$.
It is important to stress here that persisting with the $k \sim 1/t $ identification
(as done in \cite{11}) may lead to misleading conclusions about 
consistency of the cosmological solutions to the system of the equations as
given by (\ref{eq:evol}), (\ref{eq:Friedmann}), (\ref{eq:LamIR}) and 
(\ref{eq:GIR}). The observation in \cite{11} was that consistent solutions to 
the above system (with $K = +1$ or $K = -1$) exist only for a radiation 
dominated universe ($w = +1/3 $). The final conclusion in \cite{11} actuated by 
this observation was that our universe could fall into the basin of attraction 
induced by the IR fixed point only if it is spatially flat ($K=0 $). 
On the 
contrary, our scale-setting procedure shows that consistent solutions 
to the above system can be obtained for a universe having arbitrary curvature 
and for the matter equation of state with arbitrary $w$.   

The preceding paragraphs demonstrate how our scale-setting procedure leads to the mathematically consistent choice for the scale $k$ for any geometry or the matter content of the universe. Once the specification of the scale $k$ is available, one can consider its physical interpretation. The formulation of the underlying formalism, which leads to the RGE equations for $\Lambda$ and $G$, may generally put some constraints on the physically acceptable choices for the scale $k$. One such constraint is that this scale should have a geometrical interpretation
\cite{11}. This requirement is of qualitative nature and still allows for many geometrically motivated choices for $k$ \cite{85}. In finding the solution of the dynamics of the universe, one, therefore, has to obtain a solution for $k$ with an acceptable geometrical interpretation which at the same time satisfies Bianchi identity. One possible approach is first to choose an {\em Ansatz} for $k$ in such a way that it has a proper geometrical interpretation. Then, in the second step, one checks whether the set of equations describing the cosmology has a mathematically consistent solution. In the case of the mathematical inconsistency one should either discard the universe with a given geometry and the matter content as unphysical (should one decide to insist on the Ansatz) or try with some other suitable choice for $k$.
It is conceivable that in some cases of physical interest one might exhaust the list of obvious Ans\"{a}tze without finding the one which would lead to a mathematically consistent set of cosmological equations. Such an outcome certainly would not imply that the appropriate choice for $k$ does not exist for a given geometry or the matter content, but that some nontrivial choice might exist which still has a satisfactory geometrical interpretation. A different approach is based on the scale-setting procedure introduced in this paper. The scale-setting procedure always yields a mathematically consistent solution for $k$. Given this solution, it can be further examined to test whether it meets the physical requirements. This procedure is {\em systematic} and allows treatment of any geometry or the matter content of the universe.
For example, from Fig. \ref{fig:1} one can see that for the universe with a small curvature the scale $k$ obtained by our scale-setting procedure is reasonably close (at times when the IR fixed point is supposed to dominate) to the scale obtained for the flat case and it still has  a satisfactory geometrical interpretation. In this way the models with small, but nonvanishing curvatures are physically acceptable as a possible description of the universe within the RGE framework discussed in this section. This conclusion is expected since the consistency of the model for the flat space and inconsistency for the arbitrarily small $|\Omega_{K}^{0}|$ would represent a certain discontinuity and would be very surprising.       



\section{Cosmological implications}

\label{sec4}

An interesting extension of the example given in the preceding section is the 
study of a more general running of $\Lambda$ and $G$ in the infrared region.
Namely, it is interesting to consider the running in the form
\begin{equation}
\label{eq:Lamgen}
\Lambda(k) = A k^{\alpha} \, ,
\end{equation}
\begin{equation}
\label{eq:Ggen}
G(k) = \frac{B}{ k^{\beta}} \, ,
\end{equation} 
where $A$, $B$, $\alpha$ and $\beta$ are positive constants. This general
example clearly comprises the case of the IR fixed point considered in the 
preceding section, but also allows for a much more general cosmological 
evolution. The application of our scale-setting procedure to the RGE-based 
cosmology yields the following expression for the scale $k$:
\begin{equation}
\label{eq:scalgen}
k=\left( \frac{8 \pi \beta B}{\alpha A} \rho_{m} \right)^{1/(\alpha + \beta)} 
\, .
\end{equation}
The scale identified in such a manner results in the following laws of 
evolution for the quantities $\Lambda$ and $G$:
\begin{equation}
\label{eq:Lamlaw}
\Lambda = \left( \frac{8 \pi \beta}{\alpha} \right)^{\alpha/(\alpha+\beta)}
B^{\alpha/(\alpha+\beta)} A^{\beta/(\alpha+\beta)}
\rho_{m}^{\alpha/(\alpha+\beta)} \, ,
\end{equation}
\begin{equation}
\label{eq:Glaw}
G = \left( \frac{8 \pi \beta}{\alpha} \right)^{-\beta/(\alpha+\beta)}
B^{\alpha/(\alpha+\beta)} A^{\beta/(\alpha+\beta)}
\rho_{m}^{-\beta/(\alpha+\beta)} \, .
\end{equation}
These two evolution laws reveal an interesting feature of the cosmology
determined by (\ref{eq:Lamgen}) and (\ref{eq:Ggen}). Namely,
\begin{equation}
\label{eq:ratiogen}
\frac{\rho_{\Lambda}}{\rho_{m}} = 
\frac{\Lambda}{8 \pi G \rho_{m}} = \frac{\beta}{\alpha} \, .
\end{equation}
The ratio of the shares in the total energy content of the universe of the two
components is determined by two exponents $\alpha$ and $\beta$. If we consider a
scenario of the cosmological evolution in which the running of the type given by
(\ref{eq:Lamgen}) and (\ref{eq:Ggen}) sets in relatively recently (on a 
cosmological time scale) and is preceded by a long period of a very mild running 
of $\Lambda$ and $G$, then it might be possible to explain the results of 
various cosmological observations \cite{1,2} for a combination of exponents 
$\beta/\alpha \approx 2$. To verify this hypothesis, it would be necessary to 
perform an analysis similar to that made in \cite{bentivegna}.

If we combine the results (\ref{eq:Lamlaw}) and (\ref{eq:Glaw}) with
(\ref{eq:scalmatt}), we obtain the scaling laws
\begin{equation}
\label{eq:asymlaw}
\Lambda \sim G \rho_{m} \sim a^{-3(1+w) \alpha/(\alpha+\beta)} \, .
\end{equation}
The most interesting consequence of such a scaling law is the possibility that
in the distant future $\Lambda$ and $G \rho_{m}$ decrease faster than the 
curvature term, i.e. faster than $a^{-2}$. For this to happen, the following 
condition must be satisfied:
\begin{equation}
\label{eq:cond}
\frac{\beta}{\alpha} < \frac{1+3w}{2} \, .
\end{equation}
Therefore, for the parameters $\alpha$, $\beta$ and $w$ in such a relation, in
the distant future the universe becomes curvature dominated and may, in the 
case of the closed universe, cease expanding and revert to contraction. From
(\ref{eq:ratiogen}) it is clear that presently the universe cannot be described 
by the RGE-based cosmological model given by (\ref{eq:Lamgen}) and 
(\ref{eq:Ggen}), with $\alpha$ and $\beta$ such that the universe becomes 
curvature dominated in the future. However, it is possible that this sort of 
running sets in at some future instant and may lead to the curvature-dominated 
universe.


\section{RGE cosmological models from quantum field theory on curved space-time}

\label{sec5} 

As a last example, let us consider a generic case when both $\rho_{\Lambda}$ 
and $G$ can be expressed as series in the square of the RGE scale $\mu$. This 
case naturally appears in the considerations of RGE for $\rho_{\Lambda}$ and 
$G$ in the formalism of quantum field theory in curved spacetime \cite{9,10,12} 
when the influence of mass thresholds on the RGE flow is properly taken into 
account \cite{10}.

Quantum field theory on curved space-time has its motivation in effective 
field theories, especially in the very successfull chiral perturbation theory 
which is  a low-energy nonlinear realization of QCD. The framework of the
effective 
field theories enables one to renormalize otherwise nonrenormalizable
theories. Gravity, with quantum corrections being small up to the very
high Planck scale, appears to be, {\it prima facie}, an even better effective 
field theory, which is expected to govern the effects of quantum
gravity at low-energy scale without the knowledge of the true quantum
gravity. However, the reality  is not so simple, since the possible
existence of the singularity in the future, corresponding to the wavelength 
probed of the order of the size of the universe - deeply in the 
infrared region - may break the validity of the effective theory in the 
infrared limit. As a matter of fact, it was shown \cite{Tsamis}
that quantum gravity may
lead to the very strong renormalization effects at large distances induced
by infrared divergences. In the following discussions we assume the validity of the effective theory at present time and present scales persists in the
infrared region. However, we are aware that possible infrared effects in quantum
gravity may influence our conclusions in an unexpected way. It is, for example, well known \cite{Martin} that,
in nonperturbative quantum gravity, the G runs to the infinity
in the infrared limit,  which is an artifact of 
the Einstein-Hilbert
truncation, i. e., the absence of  high derivative terms.

The next 
important question is the determination and the meaning of the RGE scales. 
 In the $\bar{MS}$ scheme, the $\mu$ dependence in the effective action 
is compensated by the  running of the parameter $\Lambda$ as in QED where the $\mu$ dependence is compensated by the running charge $e(\mu)$.
The overall effective action which contains a running $\Lambda(\mu)$ is 
therefore scale independent, as required.

The physical interpretation of the RG scale in the high derivative sector can be achieved calculating \cite{Gorbar}
the polarization operator of gravitons arising from the 
particle loops in the linearized gravity. 
In the RG equation which contains the derivative with respect to the arbitray
renormalization parameter $\mu$, one usually eliminates the $\mu$ parameter
trading it for a Euclidean momentum $p$. In the physical mass-dependent
renormalization scheme (e. g. momentum subtraction scheme) $\mu$ is traded for
a certain Euclidean momentum $p$.  In QCD, for example, one writes an RG
equation with respect to the momentum scaling
parameter $\lambda$, $p\rightarrow \lambda p$, and
eliminates the derivatives with respect to $\mu$. The trade is performed by
identifying the new scale with the typical average energy (momentum) of the 
physical process. On the other hand, the use of the physical mass-dependent
renormalization scheme allows us to control the decoupling of the massive
particles in an unambigous way.

In the UV limit, the physical mass-dependent renormalization scheme will
coincide with the results obtained using the minimal subtraction scheme, but
the latter fails to describe the low momentum behavior of the beta functions.

The problem that appears is the following: as 
regards the running of $\Lambda$ and $(1/G)$ on the curved background
there is no known method of calculating that would be compatible with
some physical scheme (e. g. the momentum subtraction
scheme). One is inevitably forced to use phenomenology and an intuition if one
wants to determine the meaning of the RG scale.
Intuitively, one expects the scale to coincide with the typical momentum of
gravitinos which appear in the vertices of the Green functions. However, this 
``determination" relies heavily on our feeling of the underlying phenomenology.

In the following  we restrict our considerations to the
field theory defined
 at the classical background, the approach developed in \cite{9,10}.
Such a model was based on the observation \cite{9}
that even a ``true'' CC in
such theories cannot be fixed to any definite constant (including zero)
owing to the renormalization-group (RG) running effects. The variation of
the CC
arises solely from  particle field fluctuations, without introducing any
quintessence-like scalar fields. Particle contributions to the
RG  running of the CC which are due to vacuum fluctuations of
massive fields have been properly derived in \cite{10}, with a somewhat
peculiar outcome that more massive fields do play a dominant role in the
running at any scale.

When the RGE scale $\mu$ is less than all masses in the 
theory, we can write \cite{10,12}
\begin{equation}
\label{eq:Lamexp}
\rho_{\Lambda} = \sum_{n=0}^{\infty} C_{n} \mu^{2n} \, ,
\end{equation}
\begin{equation}
\label{eq:Gexp}
G^{-1} = \sum_{n=0}^{\infty} D_{n} \mu^{2n} \, .
\end{equation}  
We assume that both series converge well and can be well approximated by 
retaining just a first few terms. The application of the scale-setting
procedure yields the expression
\begin{equation}
\label{eq:rhommu}
\rho_{m} = -C_{0} + \frac{C_{1} D_{0}}{D_{1}} + 2 \frac{D_{0}}{D_{1}}\left(
C_{2} - \frac{C_{1} D_{2}}{D_{1}} \right) \mu^{2} + \dots \, , 
\end{equation}
which finally leads to the identification of the scale $\mu$
\begin{equation}
\label{eq:murhom}
\mu^{2} \simeq \frac{1}{2} \frac{1 - \frac{D_{1}}{C_{1} D_{0}} (\rho_{m} +
C_{0})}{\frac{D_{2}}{D_{1}}-\frac{C_{2}}{C_{1}}} \, .
\end{equation}

We can cast more light on this general expression by more precisely specifying
the values of the coefficients $C_{i}$ and $D_{i}$. From the studies of the
cosmologies with the running $\rho_{\Lambda}$ in the formalism of quantum field
theory in curved spacetime \cite{9,10,12}, we know that generally $C_{1} \sim 
m_{max}^{2}$, $C_{2} \sim N_{b}-N_{f} \sim 1$, $C_{3} \sim 1/m_{min}^{2}$, etc. 
Here $m_{max}$ and $m_{min}$ denote the largest and the smallest masses of 
(massive) fields in the theory, respectively, and $N_{b}$ and $N_{f}$ stand for 
the number of bosonic and fermionic massive degrees of freedom in the theory, 
respectively. The value of $C_{0}$ does not follow from the argumentation of 
the same type, but its value could hardly surpass $\rho_{\Lambda,0}$. Adopting 
the same line of reasoning for $G$, on dimensional grounds, we expect $D_{1} 
\sim 1$, $D_{2} \sim 1/m_{min}^{2}$, etc. For the value of $D_{0}$ it is 
natural to take $M_{Pl}^{2}$. Any other choice would imply a very strong 
running for $G$ which would not be consistent with the observational data on 
$G$ variation \cite{G}. Taking these values of the coefficients $C_{i}$ and 
$D_{i}$, we obtain the following expression for the scale $\mu$:
\begin{equation}
\label{eq:muconcrete}
\mu^{2} \simeq \frac{1}{2} \gamma_{1} m_{min}^{2} \left[ 1 - \frac{\gamma_{2}}{M_{Pl}^{2}
m_{max}^{2}} (\rho_{m} + C_{0}) \right] \, .
\end{equation}
Here $\gamma_{1}$ and $\gamma_{2}$ represent constants of order 1. This result 
for the RGE scale implies a very slow running of the scale $\mu$ in the 
vicinity of $\mu \simeq m_{min}$. This value of the scale is also only 
marginally acceptable as far as the convergence of the expansions 
(\ref{eq:Lamexp}) and (\ref{eq:Gexp}) is concerned. Furthermore, when the
solution (\ref{eq:muconcrete}) is inserted into (\ref{eq:Lamexp}), one arrives
at
$\rho_{\Lambda } \sim m_{max}^2 m_{min}^2$. If for $m_{min}$ one takes the mass 
scale revealed in recent neutrino oscillation experiments, of order $10^{-3} 
\; \mbox{\rm eV}$, one obtains $\rho_{\Lambda }^0 $, which is obviously far too 
large for any $m_{max}$. In the extreme case, we can set $m_{max} \sim M_{Pl}$ 
and saturate $\rho_{\Lambda }^0 $ with $m_{min} \sim m_{quintessence} \sim 
10^{-33} \;\mbox{\rm eV}$. Let us, however, recall that from the beginning of 
our presentation we have proclaimed quintessence degrees of freedom as 
redundant ones in the present approach. Although some intermediate scale for 
$m_{max}$ and $m_{min}$ are also possible, we do not consider this case a 
plausible one.

In the model put forward in \cite{10}, it is pointed out that some amount of 
fine-tuning needs to be introduced into the expansion (\ref{eq:Lamexp}) in 
order to maintain the consistency with the results of primordial 
nucleosynthesis. This fine-tuning is realized by the requirement $C_{1} \simeq 
0$. If we adopt this constraint, the solution for $\mu^{2}$ becomes
\begin{equation}
\label{eq:muC10}
\mu^{2} \simeq \frac{1}{2} \frac{\rho_{m} + C_{0}}{M_{Pl}^{2}} \, ,
\end{equation}
which at the present epoch of the evolution of the universe is quite well
approximated by $\mu \simeq H$, a choice advocated in \cite{12}. However, this
value of the scale leads to an extremely slow variation of $\rho_{\Lambda}$ 
with the leading term in the scale $\mu$ of the type $\mu^{4} \simeq H^{4}$.

The value of the coefficient $C_{1} \sim m_{max}^{2}$ leads to the too strong
running, while the fine-tuned value $C_{1} \simeq 0$ leads to the negligible 
running. It is natural to ask whether some intermediate value for $C_{1}$ might 
lead to a satisfactory running scheme. Assuming that $C_{1}$ has the value 
around $m_{min}$, the natural value of the scale becomes
\begin{equation}
\label{eq:C1mmin}
\mu^{2} = \gamma_{3} \left( m_{min}^{2} - \gamma_{4}\frac{\rho_{m} + C_{0}}{M_{Pl}^{2}}
\right) \, .
\end{equation}
Here $\gamma_{3}$ and $\gamma_{4}$ refer to dimensionless constants of the 
order 1. This value of the scale $\mu$ can approximately reproduce the present 
value of the cosmological term energy density $\rho_{\Lambda, 0}$ even with 
the assumption $C_{0}=0$. From (\ref{eq:C1mmin}) it is also evident that the 
scale changes very slowly. Another problem is the issue of convergence of the 
series (\ref{eq:Lamexp}) and (\ref{eq:Gexp}). However, as the ratio 
$\frac{\rho_{m} + C_{0}}{M_{Pl}^{2}}$ is at present time (and also during the 
matter-dominated era) proportional to $H^{2}$, with this choice of $C_{1}$, we 
effectively reproduce the evolution of the models elaborated in \cite{12} where 
the variability of the $\rho_{\Lambda}$ is described with the $H^{2}$ term. It 
is interesting to note that if one allows the violation of the decoupling 
theorem for $1/G$, it is possible to obtain a RGE based cosmology, consistent 
with our scale-setting procedure, in which $\mu \sim H$ and $\rho_{\Lambda}$ 
varies with $H^{2}$ \cite{H2}.

The general discussion given in this section indicates that the application of
the scale-setting procedure to the RGE cosmology determined by (\ref{eq:Lamexp})
and (\ref{eq:Gexp}) yields results at variance with observational bounds for
the values of the coefficients $C_{i}$ and $D_{i}$ determined on dimesional 
grounds only. To achieve the consistency of the model with the observational 
results, it is necessary to impose restrictions on some of these coefficients. 
The mechanism behind these restrictions, however, still remains to be clarified.

\section{Conclusions}

The procedure of the scale setting in RGE-based cosmologies presented in this
paper translates well-founded models of quantum theories to unambiguously
defined cosmological models. The underlying argument behind the entire 
procedure, the maintaining of general covariance at the level of the Einstein 
equation, is conceptually simple and theoretically convincing. The details and 
the origin of the RGE for $G$ and $\rho_{\Lambda}$ are immaterial for the 
functioning of the procedure, which is, in this respect, quite robust. In 
several examples given in this paper, the applicability of the procedure was 
demonstrated in two cases with RGE of quite different backgrounds: quantum 
gravity and quantum field theory in curved spacetime. In both cases, the 
scale-setting procedure provides insight beyond the phenomenologically 
motivated choices for the RGE running scale. With the availability of this 
scale-setting procedure, emphasis is now put on the more precise formulation of 
RGE for the cosmologically relevant quantities in the underlying fundamental 
theories.

{\bf Acknowledgements.} 
The authors would like to thank J. Sol\`{a} for useful comments on the manuscript. 
This work was supported by the Ministry of Science, Education and Sport of the 
Republic of Croatia under the contract No. 0098002 and 0098011.


\begin{thebibliography}{88}
\bibitem{1} S. Pearlmutter et al., Astrophys. J. 517  (1999) 565; A. G. Reiss 
et al., Astronom. J. 116 (1998) 1009.
\bibitem{2} C. L. Bennett et al., astro-ph/0302207; D. N. Spregel et al.,
astro-ph/0302209; H. V. P. Peiris et al., astro-ph/0302225.
\bibitem{3} A. Melchiorri, L. Mersini, C. Odman and M. Trodden, Phys. Rev.
D 68 (2003) 043509.
\bibitem{35} P. J. E. Peebles, B. Ratra, Astrophys. J. Lett., 325 L 17
(1988); C. Wetterich, Nucl. Phys. B 302 (1988) 668;
\bibitem{4} R. R. Caldwell, R. Dave, P. J. Steinhardt, Phys. Rev. Lett. 80
(1998) 1582.
\bibitem{5} P. J. Steinhardt, L. Wang and I. Zlatev, Phys. Rev. D 59 (1999)
123504.
\bibitem{6} C. Armendariz-Picon, V. Mukhanov and P. J. Steinhardt, Phys.
Rev. D 63 (2001) 103510; M. Malquarti, E. J. Copeland and A. R. Liddle, Phys.
Rev. D 68 (2003) 023512.
\bibitem{7} J. Garriga, M. Livio and A. Vilenkin, Phys. Rev. D 61 (2000)
023503; S. B. Bludman, Nucl. Phys. A 663 (2000) 865.
\bibitem{Cald} R.R. Caldwell, Phys. Lett. B 545 (2002) 23.
\bibitem{Cald2} R.R. Caldwell, M. Kamionkowski, N.N. Weinberg, Phys. Rev. Lett. 
91 (2003) 071301; H. Stefancic, Phys. Lett. B 586 (2004) 5; H. Stefancic, 
astro-ph/0311247 to appear in Phys. Lett. B; H. Stefancic, astro-ph/0312484, to
appear in Eur. Phys. J. C.
\bibitem{8} M. Reuter, Phys. Rev. D 57 (1998) 971. 
\bibitem{85} A. Bonnano and M. Reuter, Phys. Rev. D 65 (2002) 043508.
\bibitem{9} I. Shapiro and J. Sola, Phys. Lett. B 475 (2000) 236.
\bibitem{10} A. Babic, B. Guberina, R. Horvat and H. Stefancic, Phys. Rev.
D 65 (2002) 085002; B. Guberina, R. Horvat and H. Stefancic, Phys. Rev. D 67
(2003) 083001.
\bibitem{11} A. Bonnano and M. Reuter, Phys. Lett. B 527 (2002) 9.
\bibitem{bertolami} O. Bertolami, J.M. Mourao, J. Perez-Mercader, Phys. Lett. B 311 (1993) 27; O. Bertolami, J. Garcia-Bellido, Int. J. Mod. Phys. D 5 (1996) 363.
\bibitem{12} I. Shapiro and J. Sola, JHEP 0202 (2002) 006; I. Shapiro and J.
Sola, C. Espana-Bonet and P. Ruiz-Lapuente, Phys. Lett. B 574 (2003) 149; 
C. Espana-Bonet, P. Ruiz-Lapuente, I. Shapiro and J. Sola, JCAP 
0402 (2004) 006.
\bibitem{QG1} I.L. Buchbinder, S.D. Odintsov, I.L. Shapiro, Effective Action in
Quantum Gravity, IOP, Bristol, 1992.
\bibitem{QG2} J. Berges, N. Tetradis and C. Wetterich, Phys. Rept. 363 (2002)
223.
\bibitem{Odintsov} S.~Falkenberg and S.~D.~Odintsov, Int.\ J.\ Mod.\ Phys.\ A  13 (1998) 607; L.~N.~Granda and S.~D.~Odintsov, Grav.\ Cosmol.\  4 (1998) 85;  L.~N.~Granda and S.~D.~Odintsov, Phys.\ Lett.\ B 409 (1997) 206.
\bibitem{Litim} D.F. Litim, Phys. Rev. Lett. 92 (2004) 201301; D.F. Litim, Phys. Rev. D 64 (2001) 105007; F. Freire, D.F. Litim and J.M. Pawlowski, Phys. Lett. B 495 (2000) 256.
\bibitem{reuwett} M. Reuter and C. Wetterich, Nucl. Phys. B 417 (1994) 181; 
M. Reuter and C. Wetterich, Nucl. Phys. B 427 (1994) 291; 
M. Reuter and C. Wetterich, Nucl. Phys. B 506 (1997) 483.
\bibitem{polyakov} A. Polyakov, in Gravitation and Quantization, Proceedings of the Les
Houches Summer School, Les Houches, France, 1992, J. Zinn-Justin and B.
Juli\'{a} (Eds.), Les Houches Summer School Proceedings Vol. 57,
North-Holland, Amsterdam, 1995.
\bibitem{bentivegna} E. Bentivegna, A. Bonanno and M.Reuter, 
JCAP 0401 (2004) 001.
\bibitem{G} see e.g. S.G. Turyshev, J.G. Williams, K. Nordtvedt Jr., M. Shao, 
T.W. Murphy Jr., gr-qc/0311039.    
\bibitem{Tsamis} N. C. Tsamis, R. P. Woodard, Phys. Lett. B 301 (1993) 351;
Ann. Phys. (NY) 238 (1995) 1; Nucl. Phys. B474 (1996) 235; 
I. Antoniadis and E. Mottola, Phys. Rev. D 45 (1992) 2013.
\bibitem{Martin}  M. Reuter, H. Weyer, JCAP 0412 (2004) 001.
\bibitem{Gorbar} E. V. Gorbar, I. L. Shapiro, JHEP 0302 (2003) 021.
\bibitem{H2} I.L. Shapiro, J. Sola and H. Stefancic,  JCAP 0501 (2005) 012.

\end{thebibliography}
\end{document}